# Energy equilibration processes of electrons, magnons and phonons on the femtosecond timescale

J. Walowski[1], G. Müller[1], M. Djordjevic[1], M. Münzenberg[1,#], M. Kläui[2], C. A. F. Vaz[3,*], J. A. C. Bland[3]

[1]IV. Phys. Inst., Universität Göttingen, Germany,

[2]Fachbereich Physik, Universität Konstanz, Germany

[3]Cavendish Laboratory, University of Cambridge, Cambridge, CB3 0HE, UK

By means of time-resolved Kerr spectroscopy experiments we relate the energy dissipation processes on the femtosecond (electron-spin relaxation time $\tau_{el-sp}$) and nanosecond timescale (Gilbert relaxation $\tau_\alpha$) and compare the results to the first microscopic model, which was proposed by Koopmans. For both energy dissipation processes, Elliot–Yafet scattering is proposed as the dominant contributor. We controllably manipulate the energy dissipation processes by transition metal doping (Pd) and rare earth doping (Dy) of a Permalloy film and find that while a change of $\tau_\alpha$ of more than a factor two is observed, $\tau_{el-sp}$ remains constant, contrary to the predictions of the model. We explain the discrepancies by relaxation channels not considered in the original microscopic model and identify thereby the applicability of the model and possible necessary extensions to the model.

Only a detailed understanding of the excitation mechanisms in ferromagnetic films allows one to predict temporal limits of specific magnetic switching of a spin-electronic device. The excitation mechanism can be achieved using a number of methods: magnetic field pulses [1-3], relativistic electron bunches [4], resonant excitation in microwave fields [5, 6], spin-torque transfer [7], anisotropy field pulses [8, 9] and the inverse Faraday effect [10]. The last process is a very promising tool for all-optical ultrafast manipulation of magnetic storage media. An understanding of the energy dissipation processes, the "magnetic viscosity", given by the Gilbert relaxation $\tau_\alpha$ [11, 12] is key to engineering the response time in a spin electronic device. However, a microscopic understanding of the processes is lacking up to now, but is of paramount importance to manipulate the magnetization involving ultrafast magnetic processes in future magnetic devices.

Fs-pump probe techniques reveal insights into the coupling of electrons, magnons and phonons on the femtosecond timescale. With this technique the electron-spin relaxation time $\tau_{el-sp}$ that determines the demagnetization of the spin system on the ultrashort timescale can be ascertained. One expects the same energy dissipation processes to be relevant for these fs timescale spin-flip processes (localized spin-flip) and coherent precession in the 100 ps range (spin-wave vector $k = 0$), that can also be measured concurrently [13]. The mechanism that connects the electron system with the magnetic excitations may be by a single spin flip process and subsequent decay into magnons is unclear up to now. In particular the role of the spin-orbit coupling that determines the role of the phonon system is controversially discussed. One way it enters is a significant change in the band structure for spin up and spin down states resulting in the Elliot-Yafet scattering [14, 15]. Based on that mechanism, a microscopic model was recently developed by Koopmans [16] predicting the energy dissipation on the experimentally observable timescales. The model describes the processes occurring in the femtosecond range for all-optical pump-probe experiments by an extended three-temperature model [17] and relates thereby the electron-spin relaxation $\tau_{el-sp}$, the Gilbert relaxation $\tau_\alpha$ and the Curie temperature $T_C$. Alternative ideas to explain ultrafast demagnetization have also been proposed: One is the band-narrowing in the final state recently observed in fs x-ray spectroscopy [18]. The band-narrowing increases the orbital moment and thus the energy scale of the spin-orbit interaction. The idea of the quenching of the exchange interaction in the excited state is an alternative ansatz supported by experimental findings in time-resolved photoemission experiments [19]. One can think of a kind of "loose spin" model where the energy barrier for a spin-flip is significantly reduced at hot electron temperatures. It is clear that the band structure in the final state will change: a precise prediction of the contributions to the energy transfer on the 100 fs timescale would be desirable but is not available to date. The Koopmans model though neglects these effects of band-structure variation. By assuming the bands to be fixed (rigid bands), the altered electronic occupancy and the resulting electronic relaxation processes alone already result in an ultrafast demagnetization. An experimental test of the model is now necessary to develop an understanding of the microscopic processes, test the limits of validity of the model and whether further modifications of the model are warranted.

The relation between microscopic ($\tau_{el-sp}$) and the macroscopic variables ($\alpha$, $T_C$) predicted by the Koopmans model

$$\tau_{el-sp} \propto \frac{\hbar}{k_B T_C} \frac{1}{\alpha} \qquad (1)$$

will be studied in the following where $\alpha$ is the Gilbert damping related to $\tau_\alpha$ and $T_C$ the Curie temperature. The concurrent measurement of $\tau_{el-sp}$ and $\alpha$ is facilitated by the availability of femtosecond all-optical demagnetization experiments (sometimes also called all-optical ferromagnetic resonance - ao-FMR) [8, 9, 20, 21]. On the fs timescale, the electron-spin relaxation time $\tau_{el-sp}$ determines the demagnetization and the uniform precession of the magnetization (Kittel mode) can be used to determine the Gilbert damping α. To manipulate the dissipation processes we start with a permalloy (Py) film ($Fe_{80}Ni_{20}$). The damping is then increased artificially by doping the Py with different concentrations of two different materials classes: 2% Pd as a transition metal and 1% Dy and 2% Dy as a rare earth to make a comparative study between them. Then the relation of the microscopic electron-spin relaxation time $\tau_{el-sp}$ and the macroscopic Gilbert damping parameter $\alpha$ is compared for the experiment and the theory.

The ferromagnetic films have been prepared on a Si substrate by Molecular Beam Epitaxy (MBE) and co-evaporation of the Py and a doping material (Pd, Dy). The thickness of the deposited film is always 12 nm (growth rates of around 0.002 nm/s). All films are capped by a 2 nm Au film for protection against oxidation. The distribution of the doping within the films has been characterized by Rutherford (RBS) backscattering and the average doping is found to be slightly below the nominal one. The doping levels were chosen on purpose to be low. Only for low doping levels is the dopant influence on the band structure of the host Py film expected to be minor and the crystal structure of the Py film is little disturbed. Because of the different size of the rare earth atom, for very high doping concentrations amorphous growth may result which has to be avoided to allow for comparison with the undoped case [22].

In Fig. 1 a) the schematic experimental fs-pump probe experiment and in b) the experimental data are shown. The exciting pump pulse uses a fluence of $40\ mJ/cm^2$ per pulse (80 fs in width, central wavelength $\lambda = 800\ nm$). The pump fluence is modulated by a mechanical chopper. Then the Kerr rotation $\theta_K(\tau)$ is measured by a probe pulse delayed by the time $\tau$ after excitation and modulated by a photo-elastic modulator (double modulation technique [23]). A static field is applied 35° out of plane to generate a ps anisotropy field pulse large enough to start a coherent precession [20, 21]. Two corresponding data sets are shown in Fig. 1 b) for the 2% Dy and undoped Py film. The applied field $H_{ext}$ is varied from 0 – 150 mT. As the field is increased, the frequency of the precession varies. The data is analyzed by use of an harmonic function with an exponentially decreasing amplitude in order to extract the period of the precession and the decay time $\tau_\alpha$ corresponding to the Gilbert damping. A detailed study of the field dependence is necessary to determine the parameters of the film, which then enter into the calculation of the energy dissipation rate $\alpha$. The Gilbert damping $\alpha$, the decay time $\tau_\alpha$, the effective magnetic anisotropy for the thin film $K_z$ and the gyromagnetic ratio $\gamma$ are related by [20]

$$\alpha = \frac{1}{\tau_\alpha \gamma \left( H_{ext} \cos\phi - \frac{K_z}{\mu_0 M_s} + \frac{M_s}{2} \right)} \tag{2}.$$

Within the standard ansatz given by the Landau-Lifshitz-Gilbert (LLG) equation the energy dissipation rate $\alpha$ is assumed to be field-independent. In practice various dissipation processes can contribute [24-28]. When the damping is field-independent, it is called "Gilbert-like" since Gilbert assumed $\alpha$ as field independent in the LLG equation. For the undoped 12 nm thick Py film, a Gilbert damping $\alpha$ of 0.008 is found. The values obtained for the Gilbert damping are given in 1 c) for different field values. With increasing doping percentage, a systematic increase of the Gilbert damping $\alpha$ is observed. For a Dy doping of 1% (2%), the Gilbert damping increases to 0.015 (0.02). Within the field range from 60 mT to 150 mT, no dependence on $H_{ext}$ is observed (Fig. 1 c), so we conclude that the dominating dissipation channels result most likely from Gilbert-like damping. Comparison of Dy and Pd as dopant, for the same doping level, shows that, as expected, Dy results in a stronger damping [29, 30].

For the doping with a transition metal, the interaction between the damping mechanisms originates from Py 4s-Pd 5d interaction, similar to that in a s-d model [24, 31]: for the rare earth, the energy is transferred to the Dy 4f moment by a 4s-4f interaction. Because of a strong anisotropy of the 4f moments the energy is transferred to lattice excitations owing to distortion of the lattice by the precessing 4f moment.

The microscopic processes, considered as the most important contribution in the Koopmans model, are spin-flips occurring in an electron-scattering event. Since due to spin-orbit interaction the spin variable for a single electron does not commute with the Hamiltonian of the system, as a consequence a band cannot be separated into "purely" spin up and "purely" spin down. This means that the electron spin is not conserved. With scattering at so-called hot-spots, where bands are strongly intermixed, there is a high probability that the electron will spin flip after the scattering event [15]. Evaluation of the timescale where the spin-flip process occurs, can be understood as follows. If one imagines an electron moving in a band that is not spin-split and neglecting spin-orbit coupling as in a standard band-structure calculation, a scattering process looks as follows (Fig. 2 a): the electron is scattered into a new state, possibly transferring some energy and momentum to the lattice. To put it simply, within the electron lattice relaxation time the atom position moves as it "feels" the altered electron distribution in, typically, a ps timescale, the characteristic timescale of phonons. Now consider a band structure with spins and some hot-spots with spin up and down band mixing owing to spin-orbit interaction that is turned on (Fig. 2 b). If the electron is scattered into a state that is partly spin up and down at a so-called hot-spot, the spin will flip with some probability on that same timescale. This mechanism directly connects the electron-lattice relaxation in the three-temperature model with the electron-spin relaxation. The degree of coupling is given by the strength of the spin-orbit interaction. It determines the degree of spin up and spin down intermixing.

In Fig. 3 the demagnetization data on the ultrafast fs timescale for the Py samples are shown. The demagnetization process happens very quickly within the first 100-200 fs and, in contrast to the proposed model, is not significantly changed by the Dy and Pd doping. To extract the detailed relaxation times, the slopes of the curves are analyzed with an analytic solution for the three-temperature model. Since these are coupled differential equations of the first order, they can be solved by a set of exponential functions. Here we take the approximation given by Dalla Longa for a specific heat of the spin system much smaller than the specific heat of the lattice [32]. The two exponential functions mirror the demagnetization given by $\tau_{el-sp}$, the exponential function determining the decrease of the electron temperature $\tau_{el-lat}$ owing to the transfer of energy to the lattice:

$$\Delta M = \theta(t) \left( \frac{A_1 \tau_{el-sp} - A_2 \tau_{el-lat}}{\tau_{el-lat} - \tau_{el-sp}} e^{-\frac{t}{\tau_{el-sp}}} + \frac{A_1 + A_2}{\tau_{el-lat} - \tau_{el-sp}} \tau_{el-lat} e^{-\frac{t}{\tau_{el-lat}}} \right) M_0 * \Gamma(t) \qquad (3)$$

For simplification, here the heat conductance is neglected. $\theta(t)$ is the step function and $*\Gamma(t)$ stands for a convolution with the laser pulse envelope determining the temporal resolution. The demagnetization times are within 150 fs to 230 fs for all films. Taking the macroscopic Gilbert damping α extracted from the experiments at the long timescale, we plot the demagnetization times $\tau_{el-sp}$ on the ordinate and the Gilbert damping α as abscissa (Fig. 3, inset). A strong decrease of $\tau_{el-sp}$ with α is expected, according to the inverse proportionality given by Eq. 1. The functional dependence is given as a comparison with respect to the value for the pure Py film. The straight line marks the value for the undoped Py film. We find that the demagnetization times do not follow the predicted decrease with increasing damping but are only little reduced for the Pd damping and even increased for the Dy damping.

At first glance these findings seem to provide a falsification of the model. These results are supported by similar findings by the Back group for Py doped with different rare earth materials [30]. But at a closer look, the assumptions made to derive Eq. 1 are not completely valid here. In general there are different kinds of energy dissipation processes contributing to the Gilbert damping α even if one assumes only intrinsic electronic energy dissipation channels. For the low doping region the effects of extrinsic energy dissipation due to two and three magnon processes generated at defects [25-27] are generally small for high quality samples [30]. Since the Py films are an MBE-grown sample and have been prepared by co-evaporation with very low evaporation rates under ultra-high vacuum (base pressure $10^{-10}$ mbar) conditions we expect also a low extrinsic contribution. The deduction of Eq. 1 is only valid for electronic energy dissipation channels that appear at the short timescale in the same way as on the long timescale. For the rare earth doping especially this is not the case: the rare earth 4f bands are almost atomic-like and narrow in energy, far above or below the Fermi energy, and thus will not contribute at short timescales as an ultrafast relaxation channel: electronic scattering processes owing to slight hybridization between the electrons excited in the permalloy and the localized 4f are less likely (Fig. 2 c). Nevertheless, on the long timescale the energy dissipation channel is a transfer of the precessional motion to the rare earth by distortion of the lattice, responsible for the increased Gilbert damping α.

In contrast, Pd states with their 5d and 6s states will be much broader in energy and more hybridized and so the Koopmans model should be more valid. For this case we find a tendency for the demagnetization time $\tau_{el-sp}$ to decrease with the increased Gilbert damping, even though the effect is found to be less than theoretically predicted so that surprisingly also for the case of Pd the Koopmans model does not completely describe the situation. While it is clear that the model is not applicable for rare earth doping, we suggest for future investigations to focus on the damping induced by transition metals where the predictions of the Koopmans model are followed at least qualitatively. Comparing damping due to different transition metals will then reveal the reasons for the quantitative discrepancies and for which dopants the assumptions of the Koopmans model are most appropriate. Furthermore a comparison with the non-local spin current damping may be useful and first ongoing investigations have yielded promising results [34].

In summary, we manipulate the microscopic energy dissipation processes for two cases by doping Py firstly with a transition metal (Pd) and secondly with a rare earth (Dy) material. Using femtosecond all-optical pump probe experiments both the ultrafast demagnetization ($\tau_{el-sp}$) and magnetic damping ($\tau_\alpha$) were ascertained within the same experiment. We clearly show that the prediction of the Koopmans model of an increase in the magnetic energy dissipation results in a faster demagnetization time is not observed in Dy-doped permalloy. The different electronic nature of the rare earth dopant suggests a different coupling between electrons, magnons and phonons on the femtosecond timescale. For Pd doping, while not a quantitative, qualitative agreement is found. We explain that the failure of the model in terms of missing electronic relaxation channels for the rare earth doping.

**Acknowledgements**

Support by the Deutsche Forschungsgemeinschaft within the priority program SPP 1133, and the SFB 513 is gratefully acknowledged.

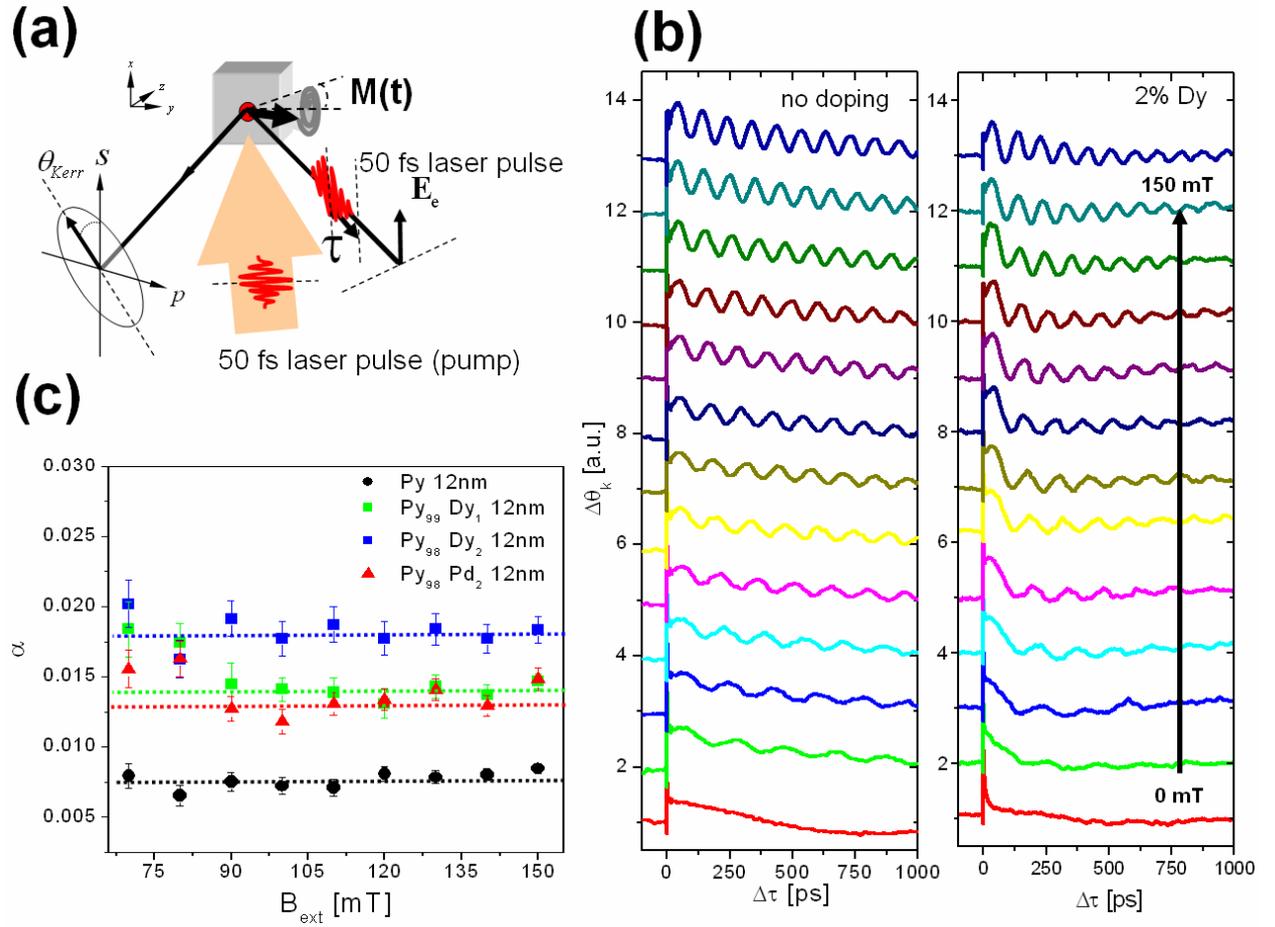

Fig. 1: (color online) (a) Schematics of the pump-probe setup. (b) Experimental data for the undoped permalloy (Py) film and with 2% Dy doping for different field values (applied 35° out-of-plane). (c) The extracted values for the Gilbert damping as a function of the applied field showing an increase with Pd and Dy doping (for details refer to the text).

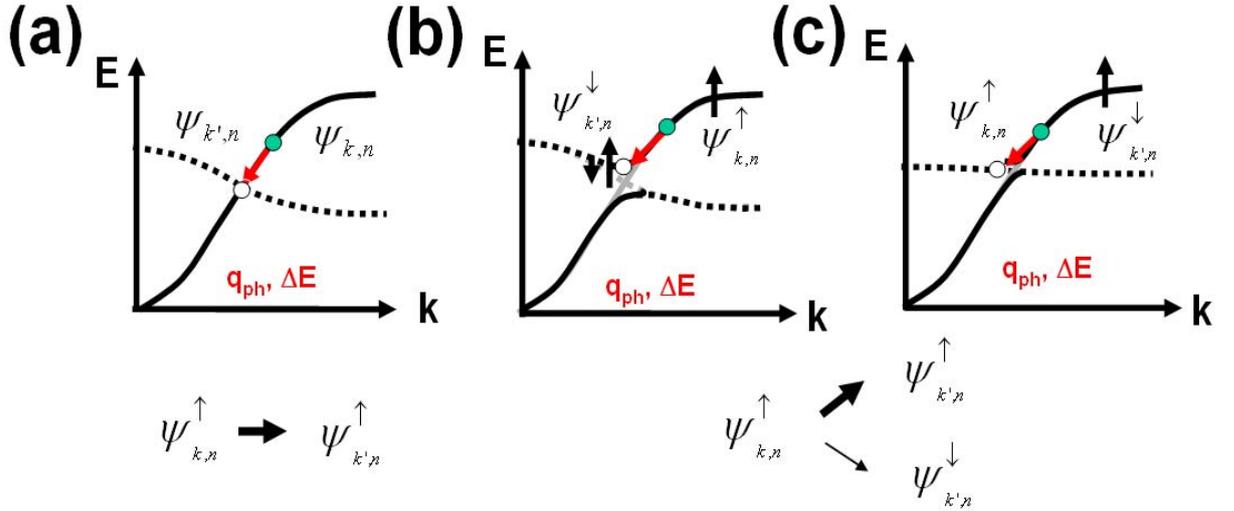

Fig. 2: (color online) The microscopic spin-flip processes leading to the spin-flips occurring in an electron-scattering event in a simplified schematic band picture. (a) The bands are not spin-split and the spin-orbit coupling is neglected. The electron is scattered into a new state in the band with no spin-flip involved. (b) Including spin-orbit interaction, the bands cannot be separated into "purely" spin up and "purely" spin down. By scattering at so-called hot-spots where bands are strongly intermixed, there is a high probability that the electron will have its spin flipped after the scattering event. (c) Rare earth 4f bands are very localized and thus only weakly hybridised.

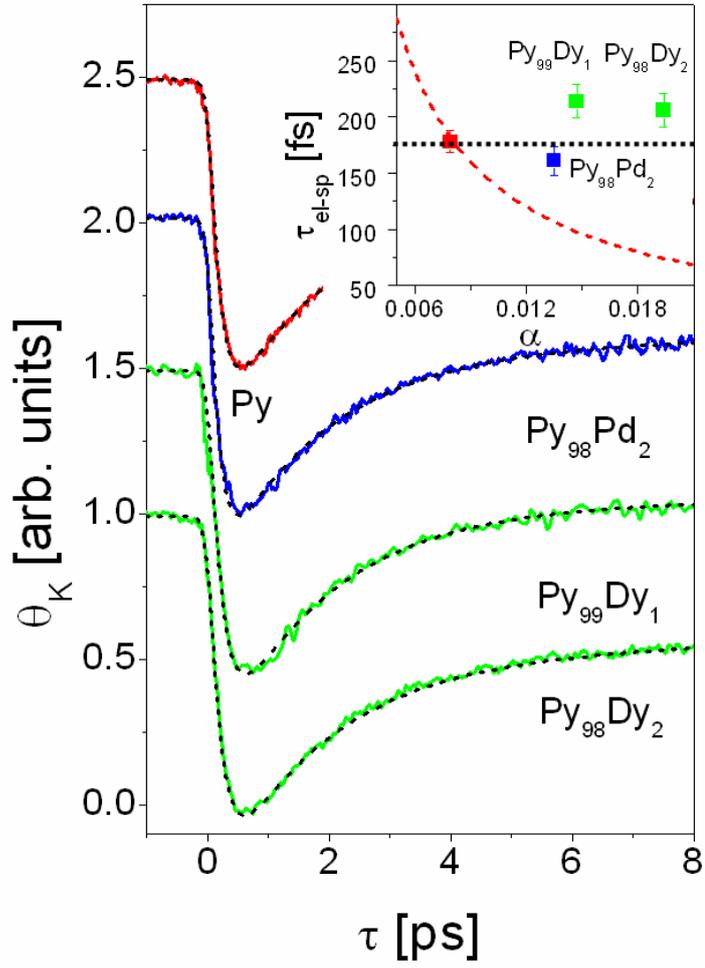

Fig. 3: (color online) The ultrafast demagnetization for 12 nm thick permalloy (Py) films with Pd (2%) and Dy (1%, 2%) doping is given. To extract the demagnetization time $\tau_{el-sp}$ is analyzed with an analytic solution of the three-temperature model. In the inset the demagnetization time is plotted as a function of the Gilbert damping α.